\documentclass[showpacs,preprintnumbers,amsmath,amssymb,prb]{revtex4}

\usepackage{graphicx}
\usepackage{bm}

\begin{document}

\title{Bose-Einstein condensation of magnons in magnets with predominant ferromagnetic interaction}

\author{A. V. Syromyatnikov}
 \email{syromyat@thd.pnpi.spb.ru}
\affiliation{Petersburg Nuclear Physics Institute, Gatchina, St.\ Petersburg 188300, Russia}

\date{\today}

\begin{abstract}
We discuss Bose-Einstein condensation of magnons (BEC) in magnets with predominant ferromagnetic (FM) interaction in magnetic field $H$ near saturation ($H_c$). Because $H_c$ is independent of FM couplings, magnetic materials of this type can have small $H_c$ that makes them promising candidates for experimental investigation of BEC. Ferromagnets with easy-plane anisotropy and antiferromagnets (AFs) containing weakly coupled FM planes or chains are discussed in detail. We observe small effective interaction between magnons near the QCP in such magnets, in contrast to AFs with strong AF coupling previously discussed. In particular, this smallness allows us to find crossovers in the critical temperature $T_c(H)\propto (H_c-H)^{1/\phi}$ from $\phi=3/2$ to $\phi=1$ in quasi-1D magnets, and from $\phi=3/2$ to $\phi\approx1$ ($T_c\ln T_c\propto H_c-H$) in quasi-2D ones.
\end{abstract}


\maketitle

\section{Introduction}

The phenomenon of Bose-Einstein condensation (BEC) is attracted much attention now. This interest is stimulated by recent achievements in experimental realization of BEC in ultracooled dilute gases \cite{atoms} and some antiferromagnets (AFs) \cite{dim1,dim2,dim3,dim4,dimorg,dim5,dim6,qcp3,dim7,af1,af2}. It has been known for a long time that a quantum spin system is equivalent to an interacting Bose gas. \cite{mm} It is demonstrated in Refs.~\cite{bat84,bat85} that this equivalence is particularly useful for discussion of AFs in magnetic field $H$ near its saturation value $H_c$. At $H>H_c$ all spins align parallel each other and excitations of the system are gapped ferromagnetic (FM) magnons. As the field becomes lower than $H_c$ the gap closes and long-range AF ordering appears in the plane perpendicular to the field that corresponds to condensation of magnons with momentum equal to AF vector. The density of condensed particles is proportional to the difference $H_c-H$ that plays the role of chemical potential. When $H_c-H\ll H_c$ one can use the well-known results for dilute Bose gas. In particular, the ground state energy and the static susceptibility were found to differ significantly from their counterparts obtained within first orders of spin-wave expansion.~\cite{bat84} The origin of these discrepancies is that interaction between magnons becomes important at $H\sim H_c$. It is properly taken into account in Bose-gas formalism within all orders of perturbation theory. Consideration of the entire perturbation series is very difficult in conventional spin-wave formalism that makes Bose-gas description to be more efficient for discussion of this problem. Then the point $H=H_c$ is a quantum critical point (QCP) that belongs to BEC universality class. When $H<H_c$ the condensate disappears at a finite temperature \cite{bat84,popov}
\begin{equation}
\label{scale}
T_c = C (H_c-H)^{1/\phi},
\end{equation}
where $\phi=3/2$ in 3D systems and $C$ is a constant. 

As the saturation field in AF is proportional to the value of the exchange coupling, $H_c$ appears to be in a hardly achievable range in the majority of AF materials. It was the main obstacle to carrying out the corresponding experiments. Only recently the first experimental observations of BEC below saturation field were reported made in Cs$_2$CuCl$_4$ ($H_c\approx8.5$ T), \cite{af1,af2} where, in particular, $\phi\approx1.5$ was observed. \cite{af1} 

In AF with singlet ground state a QCP of BEC universality class can exist in magnetic field lower than its saturation value. \cite{dim3,dim2} In this case the field larger than a gap $\Delta$ separating singlet ground state and first triplet levels generates a gas of spin-triplet states moving in a non-magnetic background. These triplets can be regarded as bosonic particles. The difference $H-\Delta$ plays the role of the chemical potential that controlls the number of particles. QCPs in a number of AFs of this type have been investigated recently both experimentally and theoretically. \cite{dim1,dim2,dim3,dim4,dimorg,dim5,dim6,dim7,qcp3} 

The possibility to produce BEC in quantum spin magnets has stimulated considerable experimental efforts to find candidate materials that have been focused on AF compounds. Then, the aim of the present paper is to draw attention to magnets with predominant FM interaction, which are equivalent near QCP to dilute Bose gas with pair interaction. Because $H_c$ is independent of FM couplings, magnetic materials of this type can have small $H_c$ that makes them promising candidates for experimental investigation of BEC. In contrast to AFs with strong AF coupling previously discussed, we observe small effective interaction between magnons in such magnets that leads to a number of particular features.

We discuss in detail two classes of such magnets. One of them contains 3D, quasi-2D and quasi-1D ferromagnets with an easy-plane anisotropy (EPF) which Hamiltonian has the form
\begin{equation}
\label{ham}
{\cal H} = -\frac12 \sum_{i,j} J_{ij} {\bf S}_i{\bf S}_j + \frac12 \sum_{i,j} A_{ij} S_i^z S_j^z + H\sum_i S_i^z,
\end{equation}
where the first term is an isotropic exchange interaction, the second one describes anisotropic exchange of the easy plane type with $z$ to be a hard axis and the last term is the Zeeman energy in transverse magnetic field $H$ directed oppositely to $z$-axis. $J$ and $A$ are implied to be positive and short-ranged. We assume below that planes in quasi-2D magnets are stacked along $z$ axis and chains in quasi-1D magnets are parallel to $z$ axis. $H_c$ is determined by the anisotropy in EPFs that is typically small being of relativistic nature. Specific calculations for EPFs similar to those for AFs \cite{bat84,bat85} have not been performed yet. Moreover the fact that QCP in EPFs belongs to BEC universality class seems to be not well recognized. To the best of our knowledge, Refs.~\cite{qcp,qcp2} are the only papers in which EPF near QCP was discussed. It was obtained for $S=1/2$ using one-loop RG approximation in the leading order of $\epsilon=2-d$, where $d$ is EPF dimension, that $H_c-H\sim T_c/|\epsilon|+{\cal O}(\epsilon^2)$ for $d>2$, i.e., $\phi=1$ in Eq.~(\ref{scale}). Meantime it remains unclear whether this formula works for $|\epsilon|\sim1$ and in particular for $\epsilon=-1$ (3D EPF). The paper~\cite{xy} should be mentioned, where $XY$ model was discussed using RG analysis which Hamiltonian is a special case of (\ref{ham}) with $A_{ij}=J_{ij}$. It was established there that for $2<d<4$ the line of phase transition in $H-T$ plane near QCP is given by Eq.~(\ref{scale}) with $\phi=d/2$. This result is consistent with our finding that the QCP belongs to BEC universality class at $d=3$. 

Another class of magnets to be considered in the present paper in detail includes quasi-2D (quasi-1D) AFs containing FM planes (chains) weakly coupled antiferromagnetically. The Hamiltonian of such systems is given by Eq.~(\ref{ham}), where $J_{ij}$ is understood to be antiferromagnetic (negative) if $i$-th and $j$-th sites belong to different planes (chains). The specific calculations for this type of AFs have not been performed yet. $H_c$ is expected to be small being determined by small AF coupling between planes (chains) and easy-plane anisotropy $A_{ij}$ (if any).

The rest of this paper is organized as follows. The Hamiltonian transformation and technique are discussed in Sec.~\ref{tec}. Effective interaction between magnons near QCP is considered in Sec.~\ref{int}. Dimensional crossover in $T_c(H)$ in quasi-low-dimensional magnets is discussed in Sec.~\ref{low}. Section~\ref{con} contains our conclusions. There are two appendixes with details of calculations.

\section{Hamiltonian transformation and technique}
\label{tec}

Let us consider first a magnet with $S=1/2$. Following Ref.~\cite{bat84} we express spin projections as 
\begin{equation}
\label{s12}
S_i^z = -\frac12 + a_i^\dagger a_i, \quad S_i^\dagger = a_i^\dagger, \quad S_i^- = a_i.
\end{equation}
where $a_i^\dagger$ and $a_i$ are Pauli operators that commute on different sites, $a_ia_i^\dagger + a_i^\dagger a_i = 1$ and $(a_i^\dagger)^2 = a_i^2 =0$. Hamiltonian (\ref{ham}) is written as
\begin{eqnarray}
\label{ham12}
{\cal H} &=& \sum_{\bf k} \left(\epsilon_{\bf k}  - \mu\right) a^\dagger_{\bf k} a_{\bf k}\nonumber\\ 
&&{}+ 
\frac{1}{2N} \sum (A_{{\bf k}_1+{\bf k}_3} - J_{{\bf k}_1+{\bf k}_3}) a^\dagger_{{\bf k}_1} a^\dagger_{{\bf k}_2} a_{-{\bf k}_3} a_{-{\bf k}_4},
\end{eqnarray}
where $\epsilon_{\bf k}  = S(J_{{\bf k}_0} - J_{\bf k})$, ${\bf k}_0$ is AF vector that is equal to $(0,0,\pi)$ and $(\pi,\pi,0)$ in quasi-2D and quasi-1D AFs, respectively, and ${\bf k}_0={\bf 0}$ in EPFs. The constant $\mu = H_c - H$ plays the role of chemical potential, where $H_c$ is the classical saturation field. $H_c = SA_{\bf 0}$ in EPFs and $H_c = SA_{\bf 0}+2S|J_{\bf 0}'|$ in AFs. The momentum conservation law ${\bf k}_1+{\bf k}_2+{\bf k}_3+{\bf k}_4=0$ is implied in the sum of the second term in Eq.~(\ref{ham12}) and in similar sums below. The bare spectrum $\epsilon_{\bf k} $ is quadratic in EPFs near ${\bf k}={\bf 0}$: $\epsilon_{\bf k}  = Dk^2$ in 3D EPF and $\epsilon_{\bf k}  = Dk_\|^2+D'k_\perp^2$ in quasi-2D (quasi-1D) EPFs, where ${\bf k}_\perp$ is the projection of $\bf k$ perpendicular to the planes (chains) and $k^2=k^2_\|+k^2_\perp$. In quasi-2D and quasi-1D AFs the spectrum is quadratic near ${\bf k}_0$: $\epsilon_{{\bf k}+{\bf k}_0}  = Dk_\|^2+D'k_\perp^2$. To treat operators $a$ as Bose ones we should introduce to the Hamiltonian (\ref{ham12}) the constraint term $U/N\sum a^\dagger_{{\bf k}_1} a^\dagger_{{\bf k}_2} a_{-{\bf k}_3} a_{-{\bf k}_4}$, where $U\to\infty$, that describes infinite repulsion of particles on the same site. \cite{bat84} As a result we lead to a gas of Bose particles with pair interaction only and with quadratic spectrum which properties are well established in the small density limit. \cite{agd,popov} In order to find physical observables one has to derive effective interaction between two particles. The ladder approximation (LA) is valid in 3D dilute Bose gas and one has to work out the equation on the vertex $\Gamma(p,0)$ shown in Fig.~\ref{energy}(a). It can be solved exactly as in Ref.~\cite{bat84}. Before discussion of the result let us consider larger spins. 

An appropriate spin transformation for $S\ge1/2$ was proposed in Ref.~\cite{bat85} basing on Holstein-Primakoff one:
\begin{subequations}
\label{ls}
\begin{eqnarray}
S_i^z &=& -S + a_i^\dagger a_i, \\ 
S_i^\dagger &=& \sqrt{2S} a_i^\dagger \left(1-\left(1-{\cal K}\right)a_i^\dagger a_i\right),  \\  
S_i^- &=& \sqrt{2S} \left(1-\left(1-{\cal K}\right)a_i^\dagger a_i\right) a_i,
\end{eqnarray}
\end{subequations}
where ${\cal K}=\sqrt{1-1/(2S)}$. Expressions for $S_i^\dagger$ and $S_i^-$ were derived from the conventional Holstein-Primakoff ones 
$
S_i^\dagger=\sqrt{2S} a_i^\dagger [1-a_i^\dagger a_i/(2S)]^{1/2}
$ 
and 
$
S_i^-=\sqrt{2S} [1-a_i^\dagger a_i/(2S)]^{1/2}a_i
$ 
by expanding the square roots, putting all operators $a_i^\dagger$ to the left of all $a_i$ using commutation relations and discarding terms containing more than three operators $a_i^\dagger$ and $a_i$ (it is reasonable because one can neglect interaction of more than two particles in dilute gas). In terms of this representation Hamiltonian (\ref{ham}) has the form (\ref{ham12}) with $S\ge1/2$ and with the additional interacting term 
$S(1-{\cal K})/N\sum (J_{\bf k_1} + J_{\bf k_3}) a^\dagger_{{\bf k}_1} a^\dagger_{{\bf k}_2} a_{-{\bf k}_3} a_{-{\bf k}_4}$. 
Equation for the vertex $\Gamma(p,0)$ shown in Fig.~\ref{energy}(a) can be solved exactly as in Ref.~\cite{bat85}. Notice that in the case of $S=1/2$ the results for the vertex obtained using the exact method discussed above and using the approximate spin representations (\ref{ls}) coincide, as in AFs considered in Ref.~\cite{bat85}. 

\section{Effective interaction between magnons near QCP}
\label{int}

Let us turn to 3D EPF first. The expression for the effective interaction $\lambda$ that is valid for $S\ge1/2$ has the form
\begin{equation}
\label{res}
\lambda_{EPF} \equiv 2\Gamma(0,0) = 
\frac{A_{\bf 0}}{1 + {\cal T}_{EPF}/(2S)},
\end{equation}
where ${\cal T}_{EPF} = 1/N \sum_{\bf k} A_{\bf k}/(J_{\bf 0}-J_{\bf k})$. Notice that $\lambda\to 0$ if $A_{ij}\to 0$ that should be contrasted to AFs with strong AF coupling, where $\lambda$ is proportional to the exchange coupling constant and ${\cal T} \sim 1$. \cite{bat84,bat85} The density of condensed particles $\rho_0$ can be found minimizing the energy shown schematically in Fig.~\ref{energy}(b) that has the form \cite{popov} 
$
E = -\mu\rho_0 + \lambda\rho_0^2/2 + 2\lambda\rho_0/N \sum_{\bf k} {\cal N}(\epsilon_{\bf k})
$, 
where ${\cal N}(\omega) = (e^{\omega/T}-1)^{-1}$. As a result one obtains \cite{popov}
\begin{equation}
\label{tc}
\rho_0 = \frac{\mu}{\lambda} - \frac2N \sum_{\bf k} {\cal N}(\epsilon_{\bf k}).
\end{equation}
Temperature fluctuations decrease $\rho_0$ that disappears at a critical temperature $T_c(H)$. After simple integration in Eq.~(\ref{tc}) we have for $T_c$ Eq.~(\ref{scale}) with $\phi=3/2$ and $C=4\pi D(2\zeta(3/2)\lambda)^{-2/3}$. \cite{popov} One finds for renormalized spectrum $\tilde\epsilon_{\bf k} = [\epsilon_{\bf k}  (\epsilon_{\bf k} +2\Lambda)]^{1/2}$ if $\Lambda\ge0$ (phase with condensate, $\rho_0\ne0$) and $\tilde\epsilon_{\bf k} = \epsilon_{\bf k}  + |\Lambda|$ if $\Lambda\le0$ (phase without condensate, $\rho_0=0$), where $\Lambda = \mu - 2\lambda/N \sum_{\bf k} {\cal N}(\epsilon_{\bf k})$. \cite{popov}  The average spin projections have the form
$\langle S^z\rangle = -S + \rho$,
$\langle S^x\rangle = \sqrt{\rho_0}\cos\psi$,
$\langle S^y\rangle = \sqrt{\rho_0}\sin\psi$,
where $\psi$ is an arbitrary phase and 
$\rho = \rho_0 + 1/N \sum_{\bf k} {\cal N}(\epsilon_{\bf k})$
is the full density of particles. Recall that all expressions for the vertex and self-energy parts are obtained within LA discarding some thermal corrections. \cite{popov} Meantime the results obtained are self-consistent except for very narrow area along the line of phase transitions, $|\Delta T|/T_c\ll \lambda\sqrt{T_c}/D^{3/2}$ (here and below $\Delta T = T_c-T$), where the omitted temperature corrections become large and should be taken into account \cite{popov} (see also Appendix~\ref{val} for discussions).

One obtains for the ground state energy $E=-\mu^2/(2\lambda)$ while its classical value is given by $E_{cl}=-\mu^2/(2A_{\bf 0})$. It is seen from Eq.~(\ref{res}) that $\lambda<A_{\bf 0}$ and quantum fluctuations lower the energy of the ground state as in AFs previously discussed. \cite{bat84,bat85} The average angle between spins and magnetic field at $T=0$ is given by $\theta \approx [2\mu/(S\lambda)]^{1/2}$ while its classical value is $\theta_{cl} \approx [2\mu/(SA_{\bf 0})]^{1/2}$. It is seen from Eq.~(\ref{res}) that quantum fluctuations renormalize strongly the ground state energy and the magnetic susceptibility if ${\cal T}/(2S)\sim1$ but one could expect ${\cal T}\ll1$ in 3D EPFs because ${\cal T} \sim A/J$. We see that in contrast to AFs with strong AF coupling effective interaction between magnons near QCP is small in EPF that leads to weak renormalization of the classical values of physical quantities. One infers that this interaction is small also in other 3D magnets with predominant FM coupling.

We cannot point out a compound that is a good 3D EPF without or sufficiently small anisotropy in the easy plane. In contrast, quasi-2D and quasi-1D materials are well-known containing FM planes or chains weakly coupling ferromagnetically (EPFs) or antiferromagnetically (AFs). We obtain for the effective interaction $\lambda_{AF}$ in the last class of compounds
\begin{equation}
\label{resaf}
\lambda_{AF} \equiv 2\Gamma(k_0,0) = \frac{A_{\bf 0} + 2|J_{\bf 0}'|}{1 + {\cal T}_{AF}/(2S)}, 
\end{equation}
where ${\cal T}_{AF} = 1/N \sum_{\bf k} (A_{\bf k} + 2J_{\bf k}')/(J_{{\bf k}_0} - J_{\bf k})$. Eq.~(\ref{resaf}) is valid for $S\ge1/2$, as is Eq.~(\ref{res}). It is seen that in contrast to EPFs the effective interaction is determined by both easy-plane anisotropy (if any) and inter-plane (-chain) AF interaction. 

It should be pointed out that the smallness of the effective interaction $\lambda\ll D$ in the magnets discussed requires to find terms proportional to $\omega$ and $\epsilon_{\bf q}$ in $\Gamma(k_0,q)$ at small $q=(\omega,{\bf q})$. Notice that these terms are negligible in AFs studied in Refs.~\cite{bat84,bat85} because $\lambda\sim D$ there. In contrast, when $\lambda\ll D$ and $T\gg\lambda$, such terms might give significant contribution for example to the last term in the expression for $E$ shown in Fig.~\ref{energy}(b). Equation for $\Gamma(p,q)$ presented in Fig.~\ref{energy}(a) can be solved exactly in quite a straightforward manner like that for $\Gamma(p,0)$ (see Appendix~\ref{vertex} for some details), the result being 
\begin{equation}
\label{gp0}
2\Gamma(k_0,q) = \lambda + (\epsilon_{\bf q} - \omega - 2\mu)\frac{(1-{\cal K})^2}{1 + {\cal T}/(2S)}.
\end{equation}
The last term here contains the {\it difference} $\epsilon_{\bf q} - \omega$ and in consequence its contribution to the last term in Fig.~\ref{energy}(b) is negligible when $\mu\ll \lambda$. 

It should be stressed that despite the formal condition of applicability of the dilute Bose-gas formalism, that is read in 3D EPF as $\rho_0^{1/3}\lambda/D\ll1$, is fulfilled for all values of $H$, our consideration is valid for $\mu\ll \lambda$ only. Really, the above formulas lead to nonsenses at $\mu\sim \lambda$. It is seen, in particular, that at $S=1/2$ Eq.~(\ref{gp0}) gives $\Gamma(k_0,0)\approx H$ that results in unrestrictedly large density of condensed particles $\rho_0=\mu/(2\Gamma(k_0,0))$ at small $H$. To explain this discrepancy we remind that dilute Bose-gas approach is applicable only when one can neglect effective interaction between more than two particles. Then, let us consider correction to the energy from diagrams containing six wavy lines that are shown in Fig.~\ref{energy}(c), where each square denotes the vertex obtained within LA. Essentially, one should take into account the dependence of the vertex on the external momenta to calculate such diagrams at $\mu\sim\lambda$ because summation over large momenta is important in this case. It is easy to realize taking into account Eq.~(\ref{gp0}) and the equality
\begin{equation}
\label{gq0}
2\Gamma(p,0) = \lambda + 2\epsilon_{\bf p}(1-{\cal K})
\end{equation}
(see Appendix~\ref{vertex} for some details of calculation of Eq.~(\ref{gq0})) that contributions from these diagrams are of the order of $D\rho_0^3$. Their sum is proportional to $\rho_0^3\Gamma^{(3)}$, where $\Gamma^{(3)}$ is the three-particles vertex with external momenta equal to $k_0$. 
\footnote{Notice that $\Gamma^{(3)}(\lambda=0)$ should be equal to zero because the point $\lambda=0$ corresponds to the isotropic ferromagnet in which all spins align parallel to a finite magnetic field, while $\Gamma^{(3)}(\lambda=0)\ne0$ would signify existence of a solution with $\rho_0\ll S$.
}
As a result one infers that intermediate processes of scattering of two particles at large momenta gives the greatest contribution to the three-particles vertex. Then we could expect that effective many-particles interaction is negligible only if $\rho_0^{1/3}\lambda_{eff}/D\ll1$, where $\lambda_{eff}\sim D$ is an effective two-particles interaction at large momenta. Thus, we lead to the same condition of applicability of the dilute Bose-gas formalism as in AFs with strong AF coupling: $\rho_0^{1/3}\ll1$. Notice that one should use effective two-particles interaction at small momenta, i.e., $\lambda$ given by Eqs.~(\ref{res}) and (\ref{resaf}), to find physical observables within LA at small $T$ and $\mu$.

The formal conditions of applicability of the dilute Bose-gas formalism in quasi-2D and quasi-1D magnets are $(\rho_0D/D')^{1/3}\lambda/D\ll1$ and $(\rho_0D^2/D^{\prime2})^{1/3}\lambda/D\ll1$ which lead at $\lambda\sim D$ to $\mu\ll D'$ and $\mu\ll D^{\prime2}/D$, respectively. At such small $\mu$ one has $T_c\ll D'$ and $T_c\ll D^{\prime2}/D$, respectively, and magnets behave like 3D one with $\phi=3/2$. LA is valid at larger $\mu$ when $\lambda\ll D$. It allows us below, in particular, to observe crossovers in $T_c(H)$ at $\mu\ll \min\{\lambda,D'\}$. Notice that the above conditions are satisfied at such $\mu$, although it is also required for this in quasi-1D magnet that $\lambda\alt\sqrt{DD'}$.

\section{Quasi-low-dimensional magnets}
\label{low}

We demonstrate now that the observed smallness of effective interaction between magnons ($\lambda\ll D$ and ${\cal T}/(2S)\ll1$) leads to  the validity of LA even at $T\gg D'$ in low-dimensional magnets. Notice that LA is not valid in AFs with strong AF exchange at $T\gg D'$. For some details about the estimations of the range of LA validity presented below (i.e., estimations of $|\Delta T|/T_c$) see Appendix~\ref{val}.
 
\subsection{Quasi-two-dimensional magnets}
\label{sec-2D} 

A magnet is considered to be quasi-two-dimensional one if $\ln (D/D')\gg1$. This criterion is fulfilled in such EPFs as K$_2$CuF$_4$ ($S=1/2$, $T_C=6.25$ K, $J=20$ K, $A=0.8$ K and $J'=1.2\cdot10^{-2}$ K) and stage-2 NiCl$_2$ graphite interlayer compound ($S=1$, $T_C=18.7$ K, $J=20$ K, $A=0.16$ K and $J'=10^{-3}$ K). \cite{lay} KCrSe$_2$ is a magnet with weak AF coupling between FM planes ($S=3/2$, $T_N=40$ K, $J=16.7$ K, and $J'\approx-0.06$ K). \cite{f-a}

\begin{figure}
\centering
\includegraphics[scale=0.4]{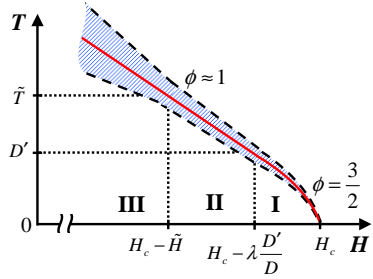}
\caption{(color online). Sketch of $H-T$ plane for quasi-2D magnet. Three regions are marked off: $\mu=H_c-H\ll\lambda D'/D$ (region I), $\tilde H\gg\mu\gg\lambda D'/D$ (region II) and $H_c \gg \mu \gg \tilde H$ (region III), where $\tilde H \sim D'$, $\tilde T \sim DD'/\lambda$ if $H_c\gg D'$, and $\tilde H \sim H_c$, $\tilde T\ll T_{C(N)}$ if $H_c\alt D'$. Solid (red) line is $T_c(H)$ that is given by Eq.~(\ref{scale}) with $\phi=3/2$ in region I and by Eq.~(\ref{tc2d}), $\phi\approx1$, in regions II and III. Dashed lines restrict the shaded (blue) area inside that the ladder approximation is not valid: $|\Delta T|/T_c\ll \sqrt{T_c/D'}\lambda/D$, $|\Delta T|/T_c\ll T_c\lambda/[DD'\ln(T_c/D')]$ and $|\Delta T|/T_c\ll 1/\ln(T_c/D')$ in regions I, II and III, respectively, where $\Delta T = T_c - T$.
\label{2d}} 
\end{figure}

When $\mu\ll\lambda D'/D\ll H_c$ (region I in Fig.~\ref{2d}) a quasi-2D magnet behaves like 3D one and the results obtained within LA are valid when $|\Delta T|/T_c\gg \sqrt{T_c/D'}\lambda/D$. In contrast to 3D EPF, summation over small momenta is important in the expressions for ${\cal T}$ and we have ${\cal T}_{EPF}=SA_{\bf 0}/(4\pi D)\ln(D/D')$ and ${\cal T}_{AF}=S(A_{\bf 0}+2|J_{\bf 0}'|)/(4\pi D)\ln(D/D')$. ${\cal T}/(2S)$ can be of the order of unity only at exponentially small $D'$ that is not the case in the materials mentioned above. After simple integration in Eq.~(\ref{tc}) we lead to Eq.~(\ref{scale}), where $\phi=3/2$ and $C=4\pi(D'D^2)^{1/3}(2\zeta(3/2)\lambda)^{-2/3}$. 

Properties of a quasi-2D magnet are more specific at $H_c\gg\mu\gg\lambda D'/D$ and $T_{C(N)}\gg T\gg D'$. Assuming that ${\cal T}/(2S)\ll1$, as in all materials mentioned above, one can replace $\Gamma(k_0,q)$ by the constant $\lambda$ in expression for $E$ shown in Fig.~\ref{energy}(b) and use Eq.~(\ref{tc}) to find $T_c$, the result being
\begin{equation}
\label{tc2d}
T_c\ln\left(\frac{T_c}{D'}\right) \approx 2\pi D\frac{\mu}{\lambda}. 
\end{equation}
Let us discuss the range of validity of LA in this case. As in 3D Bose gas, we omit thermal corrections to the vertex during the calculations. Comprehensive analysis of diagrams similar to that carried out in Ref.~\cite{popov} shows that these thermal corrections are negligible at $\tilde H\gg \mu\gg\lambda D'/D$ (region II in Fig.~\ref{2d}) when $|\Delta T|/T_c\gg T_c\lambda/[DD'\ln(T_c/D')]$, where $\tilde H\sim\min\{H_c,D'\}$. Notice that $T_c\ll DD'/\lambda$ in region II. Regions I and II exist in all compounds discussed. If $H_c\gg D'$ LA is valid also at $H_c\gg\mu\gg D'$ (region III in Fig.~\ref{2d}), but the range of validity of LA is much narrower: $|\Delta T|/T_c\gg 1/\ln(T_c/D')$. 

Region III is absent only in KCrSe$_2$ among materials listed above. Unfortunately $D'$ is too small in all these compounds and it seems to be difficult to reach the region I experimentally. Meantime Eq.~(\ref{tc2d}) can be verified in them at $T\ll T_{C(N)}$ and $\mu\ll H_c$. 

\subsection{Quasi-one-dimensional magnets}
\label{sec-1D} 

We consider a magnet to be quasi-1D one if $\sqrt{D/D'}\gg1$. This condition is fulfilled in EPF (C$_6$H$_{11}$NH$_3$)CuBr$_3$ (CHAB) ($S=1/2$, $T_C=1.5$ K, $J=110$ K, $A=5.5$ K and $J'\simeq 0.1$ K) \cite{1d} and in AF CsNiF$_3$ ($S=1$, $T_N=2.8$ K, $J=23.6$ K, $A=9.0$ K and $J'\simeq -J/500$) \cite{csnif}. 

\begin{figure}
\centering
\includegraphics[scale=0.4]{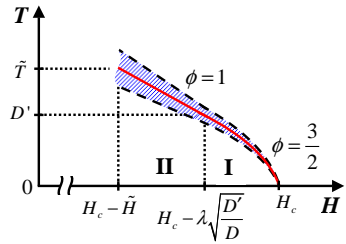}
\caption{(color online). Sketch of $H-T$ plane for quasi-1D magnet (similar to Fig.~\ref{2d} for quasi-2D), where $\tilde H \sim D'$,  $\tilde T \sim D'\sqrt{DD'}/\lambda$ if $H_c\gg D'$, and $\tilde H \sim H_c$, $\tilde T\ll T_{C(N)}$ if $H_c\alt D'$. The ladder approximation is not valid inside the shaded (blue) area determined by $|\Delta T|/T_c\ll \sqrt{T_c/D}\lambda/D'$ and $|\Delta T|/T_c\ll T_c \lambda/(D'\sqrt{DD'})$ in regions I and II, respectively. The results for region II are valid only if $A_{\bf 0}\ll\sqrt{DD'}$ in EPFs and $A_{\bf 0}+2|J_{\bf 0}'|\ll\sqrt{DD'}$ in AFs signifying the smallness of magnons interaction.
\label{1d}} 
\end{figure}

Quasi-1D magnet behaves like 3D one at $T\ll D'$ and $\mu\ll \lambda\sqrt{D'/D}\ll H_c$ (region I in Fig.~\ref{1d}). LA is valid when $|\Delta T|/T_c\gg \sqrt{T_c/D}\lambda/D'$. We have for $T_c$ Eq.~(\ref{scale}) with $\phi=3/2$ and $C=4\pi(D^{\prime 2}D)^{1/3}(2\zeta(3/2)\lambda)^{-2/3}$. One estimates ${\cal T}_{EPF}\simeq SA_{\bf 0}/(4\pi\sqrt{DD'})$ and ${\cal T}_{AF}\simeq S(A_{\bf 0}+2|J_{\bf 0}'|)/(4\pi\sqrt{DD'})$ and leads to $\lambda\simeq A_{\bf 0}/1.5$ in CHAB and $\lambda\simeq A_{\bf 0}/1.7$ in CsNiF$_3$. 

Smallness of $\lambda$ and ${\cal T}/(2S)$ allows us to use LA at $\tilde H\gg\mu\gg \lambda\sqrt{D'/D}$ (region II in Fig.~\ref{1d}) and $T_{C(N)}\gg T\gg D'$. If ${\cal T}/(2S)\ll 1$ (i.e.\ if $A_{\bf 0}\ll\sqrt{DD'}$ in EPFs and $A_{\bf 0}+2|J_{\bf 0}'|\ll\sqrt{DD'}$ in AFs) one can replace $\Gamma(k_0,q)$ by the constant $\lambda$ in Fig.~\ref{energy}(b) and lead from Eq.~(\ref{tc}) to
\begin{equation}
\label{tc1d}
T_c = \pi\sqrt{DD'} \frac{\mu}{\lambda}. 
\end{equation}
Analysis of diagrams shows that the range of validity of LA is given by $|\Delta T|/T_c\gg T_c \lambda/(D'\sqrt{DD'})$. Notice that $T_c\ll D'\sqrt{DD'}/\lambda$ in region II. Interestingly, the condition ${\cal T}/(2S)\ll1$ holds in a quasi-1D AFs with $A_{\bf 0}\alt |J_{\bf 0}'|$, where $\tilde H\sim H_c$ and LA is valid at $\mu\ll H_c$ as in 3D magnets. 

LA is not valid in the region II in the above mentioned compounds because ${\cal T}/(2S)\sim1$ in them. The exponent $\phi=3/2$ can be observed in these materials at sufficiently small temperatures $T\ll 0.1$ K and at $\mu\ll 0.5$ T.

\section{Conclusion}
\label{con} 

In summary, we discuss BEC of magnons in magnets with predominant FM interaction in magnetic field $H$ near saturation ($H_c$). EPFs and AFs containing weakly coupled FM planes or chains are discussed in detail. We find effective interactions between two particles in EPF (\ref{res}) and  in AF (\ref{resaf}). In contrast to AFs with strong AF coupling, \cite{bat84,bat85} we observe small interaction between magnons near the QCP. In particular, this smallness allows us to find crossovers in the critical temperature (\ref{scale}) from $\phi=3/2$ to $\phi=1$ (Eq.~(\ref{tc1d})) in quasi-1D magnets, and from $\phi=3/2$ to $\phi\approx1$ (Eq.~(\ref{tc2d})) in quasi-2D ones (see Figs.~\ref{2d} and \ref{1d}).

\begin{acknowledgments}

This work was supported by Russian Science Support Foundation, President of Russian Federation (grant MK-4160.2006.2), RFBR (grants 06-02-16702, 06-02-81029 and 07-02-01318) and Russian Programs "Quantum Macrophysics", "Strongly correlated electrons in semiconductors, metals, superconductors and magnetic materials" and "Neutron Research of Solids".

\end{acknowledgments}

\appendix

\section{Range of validity of the ladder approximation}
\label{val}

\begin{figure}
\centering
\includegraphics[scale=0.8]{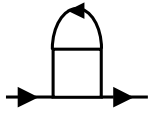}
\caption{A diagram for the normal self-energy part $\Sigma(p)$ to be taken into account in the phase without condensate. Here the square denotes the vertex obtained within the ladder approximation.
\label{la}} 
\end{figure}

We turn in this appendix to discussion of the range of validity of LA that is used for calculation in this paper. Firstly, let us consider phase without condensate. One has only normal Green's function $G(p) = 1/(\omega - \epsilon_{\bf k} + \mu - \Sigma(p))$, where $\Sigma(p)$ is normal self-energy part. $\Sigma(p)$ is determined within LA by the diagram shown in Fig.~\ref{la} that gives $2\lambda/N \sum_{\bf k} {\cal N}(\epsilon_{\bf k})$, where we use the bare Green's function $G^{(0)}(q)=1/(\omega-\epsilon_{\bf q}+\mu)$. As a result one has within LA $G^{(1)}(p) = 1/(\omega - \epsilon_{\bf k} - |\Lambda^{(1)}|)$, where 
\begin{equation}
\label{l}
\Lambda^{(1)} = \mu - \frac{2\lambda}{N} \sum_{\bf k} {\cal N}(\epsilon_{\bf k}).
\end{equation}
Recall that $\Lambda=0$ gives the line of phase transitions within LA. To find $\Sigma(p)$ in the second approximation (i.e., to obtain $\Lambda^{(2)}$) one should consider the diagram presented in Fig.~\ref{la} and use Green's function obtained within LA. As a result one has 
\begin{equation}
\label{es}
\left|\Lambda^{(2)} - \Lambda^{(1)}\right| 
\sim 
\lambda \frac{T\sqrt{\left|\Lambda^{(1)}\right|}}{D^{3/2}} \times
\left\{
\begin{array}{ll}
1, & \mbox{ 3D},\\
\displaystyle \sqrt{\frac{D}{D'}}, & \mbox{ quasi-2D},\\
\displaystyle \frac{D}{D'}, & \mbox{ quasi-1D},
\end{array}
\right.
\end{equation}
where we assume that $|\Lambda^{(1)}| \ll D'$. Expression (\ref{es}) should be much smaller than $\Lambda^{(1)}$ in order LA to be valid. For 3D systems we can represent $\Lambda$ near the line of phase transition as follows: $|\Lambda| \sim (\Delta T/T_c) \lambda(T_c/D)^{3/2}$, where $\Delta T = T - T_c$. As a result we have from Eq.~(\ref{es}) that LA is valid if $\Delta T/T_c\ll \lambda\sqrt{T_c}/D^{3/2}$. For quasi-low-dimensional systems corresponding estimations depend on the distance to the QCP (see Figs.~\ref{2d} and \ref{1d}). In quasi-2D magnets one obtains the following results. Region I: $|\Lambda| \sim (\Delta T/T_c) \lambda T_c^{3/2}/(D\sqrt{D'})$ and $\Delta T/T_c\gg \sqrt{T_c/D'}\lambda/D$. Region II: $|\Lambda| \sim (\Delta T/T_c) \lambda (T_c/D)\ln(T_c/D')$ and $\Delta T/T_c\gg T_c\lambda/[DD'\ln(T_c/D')]$. In region III one should assume that $|\Lambda| \agt D'$ that gives instead Eq.~(\ref{es}) $|\Lambda^{(2)} - \Lambda^{(1)} |\sim \lambda T/D$ and we have as a result $\Delta T/T_c\gg 1/\ln(T_c/D')$. One obtains for quasi-1D system in region I $|\Lambda| \sim (\Delta T/T_c) \lambda T_c^{3/2}/(D'\sqrt D)$ and $\Delta T/T_c\gg \sqrt{T_c/D}\lambda/D'$. In region II we have $|\Lambda| \sim (\Delta T/T_c) \lambda T_c/\sqrt{DD'}$ and $\Delta T/T_c\gg T_c\lambda/[D'\sqrt{DD'}]$.

Evaluation of the range of validity of LA in phase with condensate is somewhat more tedious task because, in particular, one has two Green's functions, normal and anomalous, at $T<T_c$. Meantime it can be done as in Ref.~\cite{popov} for 3D dilute Bose gas. The result is that LA works below the line of phase transitions as close to the line as it does above the line.

\section{Calculation of the vertex}
\label{vertex}

We present in this appendix some details of calculations of the vertex. For definiteness we discuss here EPF with $S\ge1/2$, i.e., we use spin representation (\ref{ls}). All calculations for $S=1/2$ using spin representation (\ref{s12}) can be carried out similarly and the results coincide with those for $S=1/2$ obtained using Eqs.~(\ref{ls}). AFs with predominant FM interaction can be considered in a similar way (in fact, in this case one should replace Fourier components $A_{\bf k}$ by $A_{\bf k} - 2J'_{\bf k}$ in all formulas below). It should be noted also that the method of calculation resembles that has been proposed in Refs.~\cite{bat84,bat85} to find $\Gamma(p,0)$ in AFs. 

\begin{figure}
\centering
\includegraphics[scale=0.8]{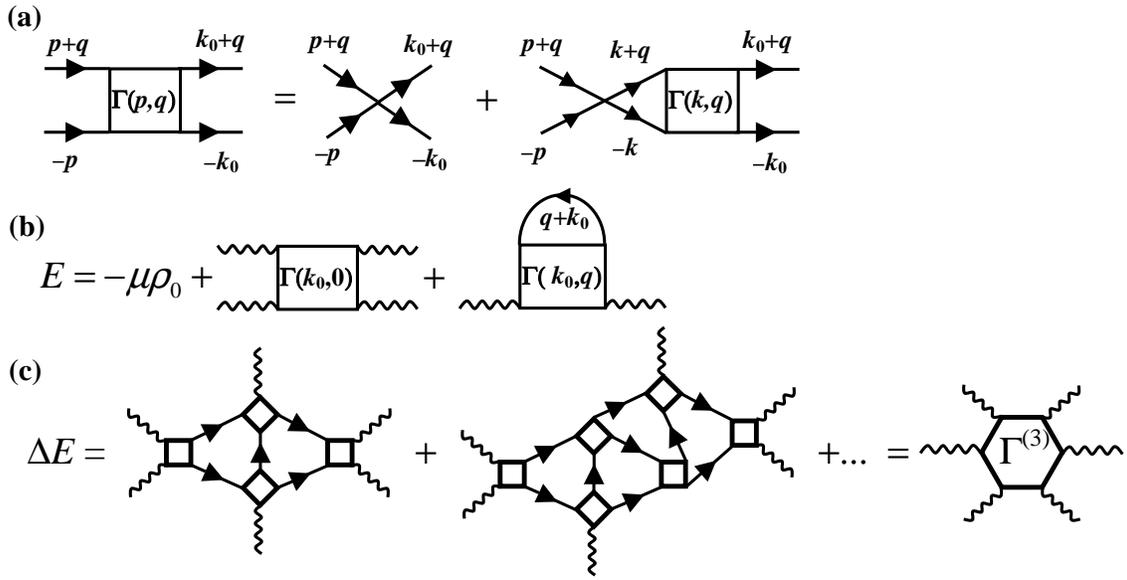}
\caption{(a) Equation for the vertex $\Gamma(p,q)$ in the ladder approximation. Here $k_0=(\omega=0,{\bf k}_0)$ and ${\bf k}_0$ is AF vector (${\bf k}_0={\bf 0}$ in EPF). (b) Part of the energy of the dilute Bose gas depending on $\rho_0$. Each wavy line denotes condensed particle and corresponds to the factor $\sqrt{\rho_0}$. (c) Correction to the energy from diagrams containing six wavy lines. Each square stands for the vertex obtained within the ladder approximation. This correction is equal to $\rho_0^3\Gamma^{(3)}$, where $\Gamma^{(3)}$ is the three-particles vertex with external momenta equal to $k_0$.
\label{energy}} 
\end{figure}

The equation for the vertex in EPF with $S\ge1/2$ that is represented graphically in Fig.~\ref{energy}(a) has the form
\begin{eqnarray}
\label{1}
4 \Gamma(p,q) &=& -J_{\bf p+q} - J_{\bf p} + A_{\bf p+q} + A_{\bf p} - 2S({\cal K} -1) (J_{\bf 0} + J_{\bf q} + J_{\bf p} + J_{\bf p+q})\nonumber\\
&&
{}+
\frac 2N \sum_{\bf k} \frac{\Gamma(k,q)}{\epsilon_{\bf k} + \epsilon_{\bf k+q} - \omega - 2\mu}
\left[
J_{\bf k-p} + J_{\bf k+p+q} - A_{\bf k-p} - A_{\bf k+p+q} + 2S({\cal K} -1) (J_{\bf k} + J_{\bf p} + J_{\bf k+q} + J_{\bf p+q})
\right],
\end{eqnarray}
where $q=(\omega,{\bf q})$, $p=({0,{\bf p}})$ and we use the bare Green's function $G^{(0)}(q)=1/(\omega-\epsilon_{\bf q}+\mu)$. Let us introduce a new quantity
\begin{equation}
\overline{\Gamma}(q) = \frac 1N \sum_{\bf p} \Gamma(p,q).
\end{equation}
Then, we have after summation over $\bf p$ of both parts of Eq.~(\ref{1})
\begin{equation}
\label{2}
4\overline{\Gamma}(q) = 2S(1 - {\cal K}) (J_{\bf 0} + J_{\bf q})
+
4S({\cal K} -1)
\frac 1N \sum_{\bf k} \frac{\Gamma(k,q)}{\epsilon_{\bf k} + \epsilon_{\bf k+q} - \omega - 2\mu}
(J_{\bf k} + J_{\bf k+q}).
\end{equation}
It is convenient to seek the solution for the vertex in the form
\begin{equation}
\label{rep}
\Gamma(p,q) = \overline{\Gamma}(q) + \alpha(q) J_{\bf p},
\end{equation}
where $\alpha(q)$ is a function independent of $p$. Substituting Eq.~(\ref{rep}) to Eqs.~(\ref{1}) and (\ref{2}) we lead to a set of two linear equations on $\overline{\Gamma}(q)$ and $\alpha(q)$ that can be readily solved. As a result one obtains, in particular, Eq.~(\ref{gp0}) for $\Gamma(k_0,q)$ and Eq.~(\ref{gq0}) for $\Gamma(p,0)$.

\bibliography{bec} 

\end{document}